\documentclass[conference,review,anonymous]{IEEEtran}
\IEEEoverridecommandlockouts
\usepackage{cite}
\usepackage{amsmath,amssymb,amsfonts}
\usepackage{algorithmic}
\usepackage{graphicx}
\usepackage{textcomp}
\usepackage{xcolor}
\usepackage{makecell}
\usepackage{multirow}
\usepackage{wrapfig,lipsum,booktabs}
\usepackage{color, colortbl}
\usepackage[utf8]{inputenc}
\usepackage[english]{babel}
\usepackage{fancyhdr}
\usepackage{lastpage}
\usepackage{caption}
\usepackage{subcaption}
\usepackage{hyperref}
\hypersetup{ hidelinks = true, }

\begin{document} 

\title{On the Lack of Consensus Among Technical Debt Detection Tools\\
}

\author{
\IEEEauthorblockN{Jason Lefever}
\IEEEauthorblockA{\textit{Drexel University} \\ jtl86@drexel.edu}
\and
\IEEEauthorblockN{Yuanfang Cai}
\IEEEauthorblockA{\textit{Drexel University} \\ yc349@drexel.edu}
\and
\IEEEauthorblockN{Humberto Cervantes}
\IEEEauthorblockA{\textit{UAM Iztapalapa} \\ hcm@xanum.uam.mx}
\and
\IEEEauthorblockN{Rick Kazman}
\IEEEauthorblockA{\textit{University of Hawai'i} \\ kazman@hawaii.edu}
\and
\IEEEauthorblockN{Hongzhou Fang}
\IEEEauthorblockA{\textit{Drexel University} \\ hf92@drexel.edu}
}

\maketitle
 
\begin{abstract}
A vigorous and growing set of technical debt analysis tools have been developed in recent years---both research tools and industrial products---such as Structure 101, SonarQube, and DV8. Each of these tools identifies problematic files using their own definitions and measures. But to what extent do these tools agree with each other in terms of the files that they identify as problematic? If the top-ranked files reported by these tools are largely consistent, then we can be confident in using any of these tools. Otherwise, a problem of accuracy arises. In this paper, we report the results of an empirical study analyzing 10 projects using multiple tools. Our results show that: 1) these tools report very different results even for the most common measures, such as size, complexity, file cycles, and package cycles. 2) These tools also differ dramatically in terms of the set of problematic files they identify, since each implements its own definitions of ``problematic''. After normalizing by size, the most problematic file sets that the tools identify barely overlap. 3) Our results show that code-based measures, other than size and complexity, do not even moderately correlate with a file's change-proneness or error-proneness. In contrast, co-change-related measures performed better. Our results suggest that, to identify files with true technical debt---those that experience excessive changes or bugs---co-change information must be considered. Code-based measures are largely ineffective at pinpointing true debt. Finally, this study reveals the  need for the community to create benchmarks and data sets to assess the accuracy of software analysis tools in terms of commonly used measures. 

\end{abstract}


\section{Introduction}
The concept of \emph{technical debt} has gained wide acceptance in the software industry \cite{ernst:techdebt21}. 
A major source of technical debt is those  \emph{not-quite-right}~\cite{cunningham:oopsla92} pieces of code that incur penalties in the form of excessive maintenance costs. Over the past few years more and more commercial and research tools have become available for developers and designers to identify \emph{not-quite-right} code. These range from well-known and widely-used commercial tools such as SonarQube~\cite{sonarsource:2013}, Structure 101~\cite{structure101}, Designite~\cite{designite}, and CAST~\cite{castsoftware}, to more recent ones such as SonarGraph~\cite{sonargraph}, CodeInspector~\cite{codeinspector}, CodeMRI~\cite{codemri}, DV8~\cite{mo:ase2018}, NDepend~\cite{ndepend}, SQuORE~\cite{squore}, SymfonyInsight~\cite{symfony}, CodeScene~\cite{codescene:codescene}, and Archinaut~\cite{cervantes:td2020}. 

Each of these tools provides its own definitions of \emph{not-quite-right} and these definitions are used to identify \emph{technical debt} candidates. For example, Designite~\cite{designite} defines \emph{smells} at architectural, design, and code levels, each based on different sets of rules. DV8~\cite{mo:ase2018,mo:tse2019} defines a suite of anti-patterns that are claimed to be design debts. Structure101~\cite{structure101} considers tangles and excessive complexity as the measures of technical debt. Most of the tools perform their analyses using source or compiled code; a few such also take change information from the revision history as input. For example, CodeScene and Archinaut detect ``hot-spots'' as files that are frequently changed, and DV8 uses co-change information in the definitions of anti-patterns. 


The existence of various ways to identify problematic files raises a fundamental question: \textit{in a given software project, will these tools report similar sets of files as being problematic?} If the answer is \textit{yes}, we can be confident that using any of these tools, the most problematic files will be identified. However, if the answer is \textit{no}, a more important question arises: \textit{which tools and which measures are more likely to report \emph{true} technical debt that incurs serious penalties that should be payed down?} These tools also provide common measures of software---such as size (lines of code), complexity, and cycles---and we should expect that they produce  similar results for these measures. 


{\color{black} To investigate these aspects we conducted an empirical study using Structure 101, SonarQube, Designite, DV8, Archinaut, and SCC. To check their consistency, we ran 14,795 files from 10 active open source projects through these tools and extracted 25  measures. We examined the correlation of each pair of measures. In addition we examined the overlaps among the sets of the 20 worst  files identified by each tool for each project.} Finally, it is well-known that the size of a file has a significant influence on various measures of concern, and so we also normalized each score by the lines of code (LOC) in each file and calculated the resulting correlations.



{\color{black} To approximate \emph{true} debt, we used four extrinsic maintainability measures extracted from each project's revision history: the number of revisions, the number of bug fixes, the number of LOC spent on revisions, and the number of LOC spent on bug fixes. These file-level measures are used as proxy measures for change-proneness, bug-proneness, and maintenance effort.} We calculated pair-wise correlations between these cost measures and each of the tool measures, using both raw and normalized scores. Our assumption is that if a tool's measure is highly correlated with revisions, bugs, and/or churn then it is likely that this measure is pinpointing files that are truly active and incurring excessive maintenance costs.

This study has produced unexpected results. First, for widely accepted and fundamental measures such as size and complexity, the tools we studied often report dramatically different numbers, meaning that the most complex files reported by one tool may not be most complex according to another tool. Package cycles and file cycles reported by different tools also differ. Second, other than size and complexity, the other measures produced by these tools barely correlate with each other;  that is, \emph{smelly} files reported by one tool are not \emph{smelly} for another. Third, except for a few measures that incorporate co-change information, the majority of these measures are not more informative than \emph{``big files are bad."}

The results suggests that, to identify files with real technical debt (that incur excessive changes or bugs), co-change (that is, historical) information should be considered. The purely code-based (that is, structural) measures, although providing insights on violations of design principles, do not appear to be effective in pinpointing real debt. {\color{black} Moreover, this study reveals the need for the technical debt community to create benchmarks and standardized data sets to assess the accuracy of software analysis tools in terms of commonly used measures. Otherwise the prospective user of a tool is always faced with an ``apples and oranges'' decision.}

\section{Research Questions}
\label{sec:rqs}

The different results provided by different analysis tools pose a challenge for practitioners who wish to select a tool and dedicate (scarce) resources to identify and fix problems in their project. To gain an understanding of this challenge, we conducted an empirical study to investigate the following research questions: 


\textbf{RQ1:} Different tools report problematic files based on their own definitions. But \emph{to what degree do these tools agree with each other by reporting similar sets of problematic files}? 

If there is a high degree of correlation among the tools, this means that any of these tools will identify a similar set of files, and we can be confident that these files are truly problematic. If there is weak or no correlation, it triggers the following questions: how do these tools differ, why do they differ, and which ones are more accurate?  

\textbf{RQ2:} For commonly recognized problems, such as cycles among files and packages, and commonly used measures, such as complexity, \emph{do these tools agree with each other}? 

If the answer is yes, we can be confident that, using any of these tools, overly complicated files and unwanted cycles won't be missed. If not, then it becomes important to ask what are the reasons behind the discrepancies? If tool $A$ identified files $a$, $b$, and $c$ as being the most complex, while tool $B$ reported files $e$, $f$, and $g$, which tool should be trusted?


\textbf{RQ3:} Our final and most important question is: do the files reported by these tools as being problematic truly contain technical debt?  That is, are the identified files associated with unreasonably high levels of effort and bugs? 

If the answer is yes, it means that the files reported as being problematic by these tools are truly debt-laden. If the answer is no, it implies that the problems identified by these tools do not pinpoint technical debt effectively.

\section{Empirical Study}
\label{sec:empiricalstudy}


To investigate these questions, our study consists of the following steps: tool and measure selection, subject selection, subject processing, and data analysis. 

\subsection{Tool selection}
Our tool selection criteria required each tool to be: 1) broadly accessible, 2) able to perform analysis at the file level, and 3) able to export results to a machine-readable, non-proprietary format. In this study, we focused on file-level analysis because a file is a minimal unit of task assignment and can be linked to maintenance costs. 

We first chose 5 well-known tools related to technical debt management that meet these criteria. Then we noticed that the size counts exported by 3 of the 5 tools did not match. So we added Succinct Code Counter (SCC)---a tool designed for counting LOC and calculating complexity---as a reference and for normalization purposes. 
For each tool we examined how they define and detect technical debt (also termed  smells or anti-patterns in these tools). Of the 6 tools, 3 of them detect cycles among files or packages, a commonly recognized design problem. Since 2 of them only report if a file participates in a cycle or not---a binary measure, their results are not comparable with the other quantitative measures. We thus separated our analysis of cycle-related measures. We now briefly introduce each tool. 

\textbf{SonarQube (SQ):} SonarQube is a widely used and well-known industrial grade automatic code review tool that defines a broad set of quality rules. For Java projects, SonarQube defines rules in 4 categories: 377 code smell rules, 128 bug rules, 59 vulnerability rules, and 33 security hotspot rules. It is not possible to analyze the measure of each individual rule as the data are too sparse to form meaningful correlation. We thus aggregated all the measures into  \emph{SQ\_Issues}, the total of SonarQube's detected rules which include \emph{CodeSmells}, \emph{Bugs}, \emph{Vulnerabilities}, and \emph{SecurityHotspots}. 
SonarQube also calculates size and complexity, but it does not provide these numbers at the file level. SonarQube takes both the source code and compiled bytecode of a project as input and displays its results in a web application.


\textbf{Designite Java (DES):} Designite
takes source code as input and identifies smells at the implementation, design and architecture levels. Because we are comparing measures at the file level, we do not consider implementation smells as they identify problems within individual methods. Designite identifies 20 types of design smells and 7 types of architecture smells. Design smells are reported at the file level, including \emph{Deep Hierarchy}, \emph{Feature Envy}, etc. The \emph{Cyclically-dependent Modularization} rule detects cycles among files. Similar to SonarQube, we created an aggregate measure, \emph{DesignSmells}, which is the sum of the reported design smells for a given file. Architecture smells are reported at the package level, including \emph{Cyclic Dependency}, which is comparable to the package cycles reported by DV8 and Structure101. We also collected size and complexity data from Designite.

\textbf{DV8:} DV8 performs the calculation of decoupling level~\cite{mo:icse2016} and propagation cost~\cite{maccormack:mnsc06}, the identification of architectural roots~\cite{xiao:icse2014}, and the identification of design anti-patterns~\cite{mo:tse2019}. Our study only considers the six anti-patterns: 1) \emph{Clique}--files forming a strongly connected component, 2) \emph{Unhealthy Inheritance}--violations of the Liskov Substitution Principle~\cite{Martin2002SOLID}, 3) \emph{Package Cycle}--two folders  mutually depend on each other, 4) \emph{Unstable Interface}--a file with many dependents that changes often with all of them, 5) \emph{Crossing}--a file with a high fan-in and fan-out that changes often with its dependents and dependees, and 6) \emph{Modularity Violation}--files that change together frequently but have no structural relationship. Similar to SonarQube and Designite, we denote the sum of all anti-patterns found in a single file as \emph{TotalIssues}. DV8 accepts the dependency information and commit history of a project as input. The dependency information is retrieved from an open source pre-processor, \textit{Depends}\cite{depends:depends},
which takes source code as input and outputs the size of each file and their dependencies in a JSON file. The commit history is retrieved by running \texttt{git-log} on a  Git
 repository with a specified date range.

\textbf{Structure101 (S101):} Structure101 is a tool for analysing, manipulating, and visualizing software architecture using the concept of a Levelized Structure Map (LSM). In addition to size, Structure101 also reports two complexity measures at the file level, which are intended to represent technical debt: \emph{Fat} and \emph{XS}. Fat is the number of edges in the dependency graph internal to a file. XS is a measure of \emph{excessive} complexity. It is a function of Fat, size, and a user provided complexity threshold. If a file's Fat exceeds the threshold, its XS will scale with its Fat and size, otherwise its XS will be zero. Structure101 also reports \emph{tangles} which are strongly connected components found in the either the package graph or class graph. We used the Studio
edition of Structure101.



\textbf{Archinaut (ARCH):} Archinaut takes various data sources as input including dependency information generated by Depends and a project's commit history. Archinaut produces various measures at the file level and can identify \emph{``hotspots''} by analyzing trends in the measures. In particular, for each file, it calculates \emph{FanIn}, \emph{FanOut}, \emph{TotalDeps} (the sum of \emph{FanIn} and \emph{FanOut}) as well as \emph{CoChangePartners} (the number of other files the target file has co-changed with, as recorded in the project's commit history.) In addition, Archinaut is capable of integrating results from other tools. In this study Archinaut is used both as a producer of measures and as an integrator of results from the other tools. It compiles the results into a single CSV file.

\textbf{Succinct Code Counter (SCC):} Succinct Code Counter is a tool for measuring code size and complexity.
\begin{table}[htbp]
\caption{Tools, Measures, and Abbreviations}
\begin{center}
\begin{tabular}{|l|l|c|c|} 
\hline
Tool:Measure &Abbr. &Aggr.\\ \hline
\multicolumn{3}{|c|}{File Measures} \\ \hline
ARCH:Dependent Partners &ARCH\_fanIn &\\ \hline
ARCH:Depends on Partners &ARCH\_fanOut &\\ \hline
ARCH:Total Dependencies &ARCH\_deps &x\\ \hline
ARCH:CoChange Partners &ARCH\_coCh(h) &\\ \hline
DES:Size &DES\_size &\\ \hline
DES:Complexity &DES\_comp. &\\ \hline
DES:Design Smells &DES\_smells &x\\ \hline
DV8:LOC &DV8\_size &\\ \hline
DV8:TotalIssues &DV8\_issues &x\\ \hline
DV8:Crossing &DV8\_crossing(h) &\\ \hline
DV8:ModularityViolation &DV8\_MV(h) &\\ \hline
DV8:UnhealthyInheritance &DV8\_UH &\\ \hline
DV8:UnstableInterface &DV8\_UI(h) &\\ \hline
SQ:Issues &SQ\_issues &x\\ \hline
SQ:CodeSmells &SQ\_smells &\\ \hline
SQ:Bugs &SQ\_bugs &\\ \hline
SQ:Vulnerabilities &SQ\_vuls &\\ \hline
SQ:SecuritySpots &SQ\_sec &\\ \hline
S101:Size &S101\_size &\\ \hline
S101:Fat &S101\_fat &\\ \hline
S101:XS &S101\_xs &\\ \hline
SCC:CLOC &SCC\_size &\\ \hline
SCC:COMPLEXITY &SCC\_comp. &\\ \hline
\multicolumn{3}{|c|}{Cycle Measures} \\ \hline
S101:ClassTangles &S101\_fileCycle &\\ \hline
S101:PkgTangles &S101\_pkgCycle &\\ \hline
DV8:Clique &DV8\_fileCycle &\\ \hline
DV8:PackageCycle &DV8\_pkgCycle &\\ \hline
DES:Cyclically-dependent Modularization &DES\_fileCycle &\\ \hline
DES:Cyclic Dependency &DES\_pkgCycle &\\ \hline
\end{tabular}
\end{center}
\label{tbl:tool-abbr}
\end{table}

Table~\ref{tbl:tool-abbr} provides a summary of the measures selected for this study. The 1st column lists each tool and its original measures. The 2nd column shows the abbreviation of these measures used for the rest of this paper. If a measure incorporates information from a project's commit history, such as co-change or the number of commits, we place a \emph{``(h)"} next to its abbreviation. The 3rd column indicates if a measure is an aggregation of other measures.

\subsection{Subject Selection}
To be selected for analysis each project has to meet several constraints. First, the project has to have at least 1000 commits that are linked to an issue tracked by an issue tracker. These links allow us to identify bug-fixing commits so we can calculate the bug churn and bug commits for each file. Second, for the project to be compatible with all of the chosen analysis tools, the project has to contain primarily Java source code.

To find an initial set of candidate projects, we ran a query against the 20-MAD dataset introduced by Claes et al. ~\cite{claes:msr20} to find the projects with the highest ratio of linked to unlinked commits. We chose a minimum of 1000 linked commits and a minimum ratio of 70\% linked to unlinked commits. Furthermore, if GitHub did not identify the project as being composed of more than 80\% Java source code, we excluded the project.

This process gave us 20 candidate projects. We further filtered this list using more practical considerations. First, both SonarQube and Structure101 run their analysis against compiled Java bytecode, not source code. This meant that we had to be able to  build each project from source. Second, some tools can not analyze large projects. We excluded these projects that cannot be analyzed on our hardware. These considerations left us with 10 projects, briefed in Table~\ref{tbl:subjects}, each of which is maintained by the Apache Software Foundation. The table also includes information about the version that was analyzed (which corresponds to the final date in the analysis period), the physical size of its Java code as measured using SCC (``$LOC$"), the number of Java files (``$\#File$"), the total number of commits (of which exactly 1000 are linked to an issue)(``$\#Com.$"), and the analysis period (``$Period$"), determined by the dates of the commits. 

\begin{table}[htbp]
\caption{Subject Projects}
\footnotesize
\begin{center}
\begin{tabular}{|l|c|c|c|c|c|}
\hline
\textbf{Project} & \textbf{Version} & \textbf{LOC} & \textbf{\#File} & \textbf{\#Com.} & \textbf{Period} \\ \hline
DeltaSpike & 1.9.2  & 147125 & 1824 & 1266 & \makecell{2014/03/16- \\ 2019/12/13}          \\ \hline
Flume & 1.9.0-rc3 & 80266 & 441 & 1082 & \makecell{2012/02/14- \\ 2018/12/17} \\ \hline
HBase & 1.4.12  & 439093 & 1786 & 1088 & \makecell{2017/01/04- \\ 2019/11/20} \\ \hline
Knox   & 1.3.0 & 82549 & 863 & 1098 & \makecell{2016/01/30 - \\ 2019/06/29} \\ \hline
NiFi & 1.10.0 & 629298 & 4112 & 1051 & \makecell{2018/05/03- \\ 2019/10/28} \\ \hline
Oozie & 5.2.0 & 151221 & 812 & 1059 & \makecell{2014/05/20- \\ 2019/10/29} \\ \hline
Qpid-B\footnotemark[1] & 7.1.6 & 321853 & 2000 & 1148 & \makecell{2017-04-17 \\ 2019/12/02} \\ \hline
Qpid-J\footnotemark[1] & 6.3.4 & 104181 & 561 & 1139 & \makecell{2016/01/25- \\ 2019/05/14} \\ \hline
Tajo & 0.12.0-rc0 & 265253 & 1557 & 1095 & \makecell{2014/03/11- \\ 2019/09/11} \\ \hline
TEZ & 0.9.2 & 159251 & 839 & 1082 & \makecell{2015/05/18- \\ 2019/03/19} \\ \hline
\end{tabular}
\end{center}
\label{tbl:subjects}
\end{table}

\footnotetext[1]{The full names of these two projects are Qpid Broker-J and Qpid JMS AMQP 0-x, respectively.}


\subsection{Subject Processing}

\begin{figure}[]
\includegraphics[width=0.49\textwidth]{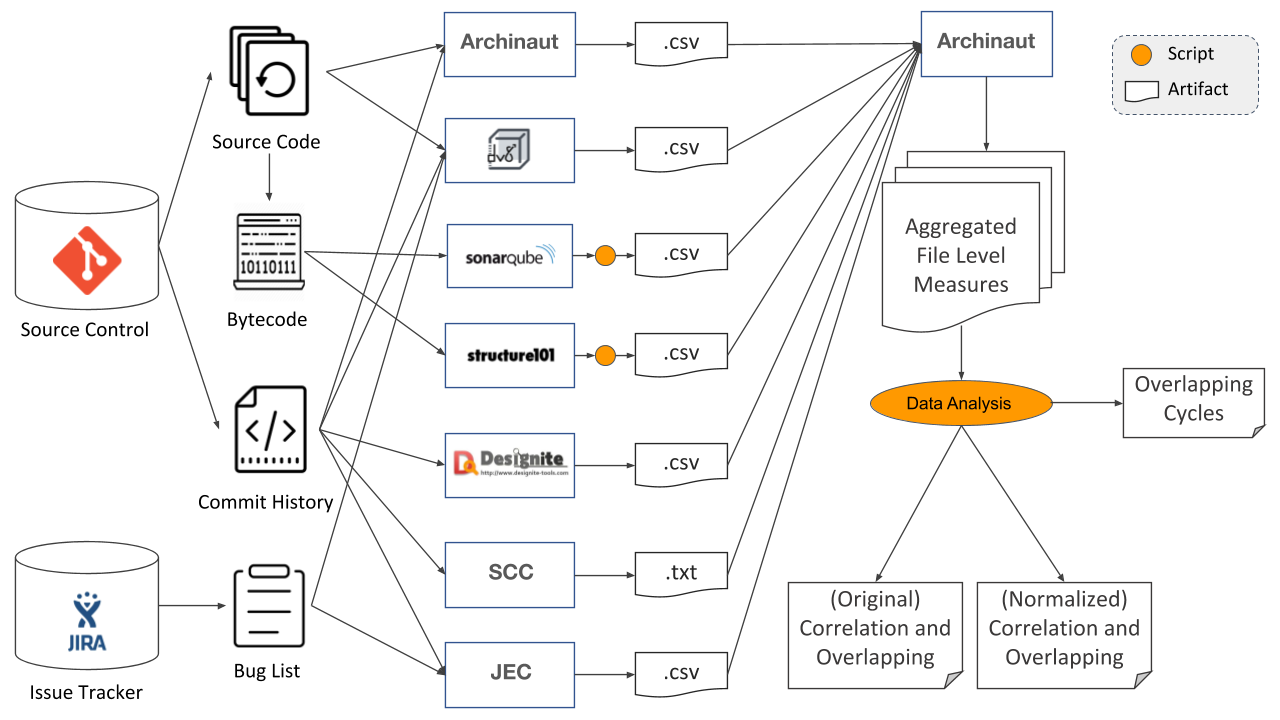}
\caption{Data Processing Overview}
\label{figure:analysisprocess}
\vspace{-0.5 cm}
\end{figure}

To prepare the input for these tools, we cloned each project repository, checked out the indicated version, and extracted its commit history. Files that contain only tests, examples, or generated code were removed. A list of issues identified as bugs in the project's issue tracker were collected into a single file. Finally, each project was built according to its provided documentation.

Then each analysis tool was run: Designite and SCC were given the raw source code; SonarQube and Structure101 were also supplied with the compiled bytecode; Archinaut and DV8 were given the commit history, the list of bugs, and the structural information generated by the Depends pre-processor.
This allowed us to extract file level results from each tool. 

Finally, for each file in each project, we collected four maintainability measures over the chosen analysis period: 
\emph{\#Change}---the number of times a file appears in commits, \emph{\#Bug}---the number of file revisions linked to bug commits, \emph{\#ChangeChurn}---the added and removed LOC in these commits, and \emph{\#BugChurn}---the added and removed LOC in file bug commits. These maintainability measures where calculated with both DV8 and our independently created tool, JEC.
This entire process is illustrated in Figure~\ref{figure:analysisprocess}.

\subsection{Data Analysis}
In our analysis, file measures and cycle data were separated. For each pair of file measures, we analyzed their relationship from two perspectives. First, we looked at their correlation. For example, is it true that error-proneness scales with code smells? We calculated the Pearson correlation coefficient using both the raw numbers and the numbers normalized by LOC (as reported by SCC). Second, we looked at the overlap of the 20 worst files identified by each measure. For example, are the smelliest files reported by Designite also the smelliest files from SonarQube? 

For each pair of cycles reported, we identified their overlapping files. This allowed us to determine whether any pair of cycles contained the same members and whether these tools detected the same set of cycles. Cycles are compared at both the file and package level. 


\section{Results}
In this section, we discuss the results of our empirical study to answer the three research questions. 


\subsection{RQ1: Do the tools agree with each other?}
\label{sec:q1}

To answer this question, we present the results for the file-based  measures and the cycle-related measures respectively. 


\textbf{a. Comparing file measures.} We compare each pair of file measures in two ways, \emph{correlation} and \emph{top 20 overlaps}. 
As size and complexity are simple, fundamental measures of code, we present their correlations separately. 

\begin{table}[htbp]
\caption{Size Measure Pair-Wise Comparison} 
\centering
\footnotesize{Avg. \#Top-20-Overlaps (above diagonal)\\ 
Avg. Correlation (below diagonal)}
\begin{center}
\begin{tabular}{|l|c|c|c|c|c|} 
\hline
Names &id &(1) &(2) &(3) &(4)\\ \hline
DV8\_size &(1) &- &17.2 &16.8 &19.9\\ \hline
DES\_size &(2) &0.98 &- &15.7 &17.1\\ \hline
S101\_size &(3) &0.97 &0.95 &- &16.7\\ \hline
SCC\_size &(4) &1.00 &0.98 &0.97 &-\\ \hline
\end{tabular}
\end{center}
\vspace{-0.3 cm}
\label{tbl:q2-size-pairs}
\end{table}

1) \textit{Correlations:} The Pearson correlation coefficients for each pair of file measures over all the files of all 10 projects were calculated. Using the size measures as an example, 4 tools export ``\emph{size}" measures for each file, forming 6 cross-tool pairs. Table~\ref{tbl:q2-size-pairs} displays the average correlation coefficients between each pair of size measures over 10 projects in the cells below the diagonal. For example, row(4), column(1) contains ``$1.00$", meaning that SCC and DV8 agree on the size of files. Strikingly, even for size measures, these tools do not fully agree with each other, as we will discuss in Section~\ref{sec:q2}.

2) \textit{\#Top-20-Overlaps:} This represents the number of shared files among the top 20 files identified by the pair of measures. This helps to compare whether two tools rank the same set of files as being problematic. The numbers \textit{above} the diagonal in Table~\ref{tbl:q2-size-pairs} are the average number of overlapping files among the top 20 for each pair. For example, row(1), column(4) contains ``$19.91$", meaning that the 20 largest files reported by DV8 are mostly the same as those reported by SCC. 

\begin{table}[htbp]
\caption{Complexity Measure Comparison} 
\centering
\footnotesize{Avg. \#Top-20-Overlaps (above diagonal)\\ 
Avg. Correlation (below diagonal)}
\begin{center}
\begin{tabular}{|l|c|c|c|c|c|} 
\hline
Names &id &(1) &(2) &(3) &(4)\\ \hline
DES\_comp. &(1) &- &13.5 &12.7 &12.8\\ \hline
SCC\_comp. &(2) &0.91 &- &10.6 &10.7\\ \hline
S101\_fat &(3) &0.66 &0.61 &- &18.7\\ \hline
S101\_xs &(4) &0.61 &0.56 &0.92 &-\\ \hline
\end{tabular}
\end{center}
\label{tbl:q2-complexity-pairs}
\vspace{-0.3 cm}
\end{table}

\begin{table}[htbp]
\caption{Normalized Complexity Measure Comparison} 
\centering
\footnotesize{Avg. \#Top-20-Overlaps (above diagonal)\\ 
Avg. Correlation (below diagonal)}

\begin{center}
\begin{tabular}{|l|c|c|c|c|c|} 
\hline
Names &id &(1) &(2) &(3) &(4)\\ \hline
DES\_comp. &(1) &- &2.9 &1.0 &1.0\\ \hline
SCC\_comp. &(2) &0.50 &- &1.3 &1.4\\ \hline
S101\_fat &(3) &0.13 &0.15 &- &15.6\\ \hline
S101\_xs &(4) &0.11 &0.14 &0.77 &-\\ \hline
\end{tabular}
\end{center}
\label{tbl:q2-norm-complexity-pairs}
\end{table}

It is well known that size has a significant impact on almost every other measure. To assess the relation among these measures {\em without} the influence of size, we also calculated the measure-pair correlations and top-20-overlaps after normalizing by LOC (using the value from SCC). Table~\ref{tbl:q2-complexity-pairs} presents the original correlation and the number of top 20 overlaps between complexity measures, and Table~\ref{tbl:q2-norm-complexity-pairs} presents the data after normalization. 

It is clear that, without normalization, the complexity measures have relatively high correlation, but after normalization, both the correlation and number of top 20 overlaps dropped significantly, except for S101\_fat and S101\_xs as the latter is a function of the former. The implication is that complexity measures are highly correlated with size measures, but their correlations among each other are weak. In other words, these tools frequently disagree on which files are most complex. 

\begin{table*}
\caption{\#Projects where Measures Agree}

\centering\footnotesize{\#Top-20-Overlaps $\geq$ 10 (above diagonal);
Correlation $\geq$ 0.5 (below diagonal)}

\footnotesize
\begin{center}
\begin{tabular}{|l|c|c|c|c|c|c|c|c|c|c|c|c|c|c|c|c|} 
\hline
Names &id &(1) &(2) & (3) & (4) &(5) &(6) &(7) &(8) &(9) &(10) &(11) &(12) &(13) &(14) &(15)\\ \hline
DES\_comp. &(1) &- &\cellcolor{lightgray}9 &\cellcolor{lightgray}9 &\cellcolor{lightgray}9 &2 & & & &2 & & &2 & & &\cellcolor{lightgray}7\\ \hline
SCC\_comp. &(2) &\cellcolor{lightgray}10 &- &\cellcolor{lightgray}6 &\cellcolor{lightgray}6 &1 & & & &3 & & &2 & &1 &\cellcolor{lightgray}5\\ \hline
S101\_fat &(3) &\cellcolor{lightgray}9 &\cellcolor{lightgray}7 &- & &1 & & & &3 & & &2 & & &3\\ \hline
S101\_xs &(4) &\cellcolor{lightgray}10 &\cellcolor{lightgray}8 & &- &1 & & & &3 & & &2 & & &4\\ \hline
ARCH\_fanOut &(5) &\cellcolor{lightgray}6 &\cellcolor{lightgray}8 &3 &2 &- & & & &\cellcolor{lightgray}7 & &1 &\cellcolor{lightgray}8 & &1 &2\\ \hline
ARCH\_deps &(6) &1 & & & & &- & & &2 & & & & & &\\ \hline
ARCH\_coCh &(7) &3 &1 & & & & &- & &3 &1 &1 & &1 &1 &\\ \hline
DES\_smells &(8) & & & & &1 & & &- & & & & & & &1\\ \hline
DV8\_issues &(9) &\cellcolor{lightgray}7 &\cellcolor{lightgray}7 &3 &1 &8 &4 &\cellcolor{lightgray}5 & &- & & & & & &2\\ \hline
DV8\_crossing &(10) &1 & & & &1 & &2 & & &- & & & & &\\ \hline
DV8\_MV &(11) &2 &1 & & &3 & &\cellcolor{lightgray}6 & & & &- & & & &\\ \hline
DV8\_UH &(12) &\cellcolor{lightgray}5 &\cellcolor{lightgray}5 &4 &4 &\cellcolor{lightgray}8 &1 &2 & & & & &- & & &1\\ \hline
DV8\_UI &(13) &1 &1 &1 & &2 &1 &3 & & & & & &- & &\\ \hline
SQ\_issues &(15) &\cellcolor{lightgray}10 &\cellcolor{lightgray}10 &\cellcolor{lightgray}5 &\cellcolor{lightgray}6 &\cellcolor{lightgray}6 &1 & &1 &2 & & &3 &1 & &-\\ \hline
\end{tabular}
\end{center}
\label{tbl:q1-pairs}
\end{table*}

\begin{table}[htbp]
\caption{\#Projects where Normalized Measures Agree}

\centering\footnotesize{\#Top-20-Overlaps $\geq$ 10 (above diagonal) \\
Correlation $\geq$ 0.5 (below diagonal)}

\begin{center}
\begin{tabular}{|l|c|c|c|c|c|c|c|c|c|c|} 
\hline
Names &id &(1) &(2) & (3) & (4) &(5) &(6) &(7) &(8)\\ \hline
DES\_comp. &(1) &- & & & & & & &\\ \hline
SCC\_comp. &(2) &4 &- & & & & & &\\ \hline
ARCH\_deps &(3) & & &- & & & & &\\ \hline
ARCH\_coCh &(4) & & & &- & & & &\\ \hline
DV8\_issues &(5) & & &1 & &- & & &\\ \hline
DV8\_MV &(6) & & & &2 & &- & &\\ \hline
DV8\_UI &(7) & & & &1 & & &- &1\\ \hline
SQ\_issues &(8) & & & & & & & &-\\ \hline
\end{tabular}
\end{center}
\label{tbl:q1-norm-pairs}
\end{table}

Table~\ref{tbl:q1-pairs} and \ref{tbl:q1-norm-pairs} present the relations among non-size measure-pairs in a slightly different way. Since most of the calculated correlations are very low, and average correlations over all 10 projects are too small to be informative, we chose to present (in the lower triangle of the tables) the number of projects in which a pair of measures has a correlation greater than 0.5, the threshold that indicates a moderate correlation. For example, row(15), column(1) contains ``$10$", meaning that for all 10 projects, SQ\_issues exhibits a positive correlation ($\geq 0.5$) with DES\_complexity. 

The numbers {\em above} the diagonal show the number of times the top 20 files from each pair have more than 10 overlaps (that is, at least 50\% of the top 20 files match). For example, row(1), column(15) contains 7, meaning that in 7 out of the 10 projects, DES\_complexity and SQ\_issues share at least 10 files among their 20 top ranked files. 

Note that row(15), column(8) contains ``$1$", meaning that for SQ\_issues vs. DES\_smells, only in 1 in 10 projects do their top 20 picks share 10 or more files. The implication is that the files with design smells as detected by Designite are mostly different than those detected by SonarQube. 

Table~\ref{tbl:q1-norm-pairs} presents the same data after normalization by size. The table shows that except for the two complexity measures, only three pairs of measures have a correlation greater than 0.5 in 1 or 2 projects: DV8\_issues vs. ARCH\_deps (row(5), column(3)), DV8\_MV vs. ARCH\_coCh (row(6), column(4)), and DV8\_UI vs. ARCH\_coCh (row(7), column(4)). The top 20 overlapping file data are even more sparse: only 1 pair, DV8\_UI and SQ\_issues, shared more than 10 files in 1 project among their top 20 picks (row(7), column(8)). Note that after normalization, in no project were the most complex files identified by S101 overlapping with 10 or more files identified by the other complexity measures. 

The implication is that the largest and most complicated files often have other design issues. After normalizing, to remove the impact of size, in the majority of cases the most problematic files identified by one tool are very different from those identified by the other tools. These consistent discrepancies raises the question of which tool can be relied upon to identify the truly problematic files. 

\begin{table*}[htbp]
\caption{Package and File Cycle Comparison}
\begin{center}
\begin{tabular}{|l|c|c|c|c|c||c|c|c|c|c|c|c|c|c|} 
\hline
\multirow{3}{*}{Project}&\multicolumn{5}{c|}{Package Cycle Comparison} & \multicolumn{5}{c|}{File Cycle Comparison}\\
\cline{2-6} \cline{7-11}
&\multicolumn{2}{c|}{DES-DV8-S101} &DES $\cap$ & DES $\cap$ &DV8 $\cap$ &\multicolumn{2}{c|}{DES-DV8-S101} & DES $\cap$ & DES $\cap$ &DV8$\cap$ \\
\cline{2-3} \cline{7-8}
&\#set & \#pkg &DV8&S101&S101 &\#sets &\#file & DV8 & S101 & S101\\
\hline
DeltaSpike& 2-29-9 & 8-34-44 & 7 & 8 & 34 & 1-9-12 & 15-33-45 & 10 & 15 & 32\\ \hline
Flume& 1-12-5   & 3-18-20 & 3  & 3  & 18 & 1-8-11  & 25-47-47   & 21 & 21 & 43\\ \hline
HBase&10-122-10 &52-71-86 & 36 & 37 & 68 & 1-36-40 & 371-521-656 & 309 & 348 & 495\\ \hline
Knox&4-32-10    &7-40-50 &  5& 4& 36 & 1-20-15 &57-82-50& 43 & 21 & 32\\ \hline
NiFi& 16-145-37 &41-157-164 & 36 & 34 & 141 & 1-86-62 & 144-330-354 & 90 & 118 & 217\\ \hline
Oozie& 3-76-4 & 28-44-45 & 28 & 27 & 41 & 1-13-20 & 168-389-253 & 159 & 111 & 237\\ \hline
Qpid-B& 8-117-8 &15-82-97 & 15 & 15 & 78 & 1-54-41 & 263-332-586 & 190 & 236 & 283\\ \hline
Qpid-J& 3-23-4 &12-23-29 & 11 & 12 & 23 & 1-7-8 & 263-332-586 & 147 & 170 & 205\\ \hline
Tajo&   8-77-11 & 37-68-75 & 37 & 37 & 68 & 1-33-39 & 182-452-527 & 160 & 174 & 442\\ \hline
TEZ& 5-67-10 & 26-56-65 & 25 & 26 & 54 & 1-17-17 & 79-136-155 & 56 & 69 & 113\\ \hline
\end{tabular}
\end{center}
\label{tbl:q2-all-cycles}
\end{table*}

\textbf{b. Comparing cycle measures.} Three tools we studied were able to detect cycles at both file and package levels. As cycles are relatively straightforward to define we expected that these tools would report consistent results. This was not the case, as shown in  Table~\ref{tbl:q2-all-cycles}. 

For example, the 1st row in the ``\emph{Package Cycle Comparison}" section shows that in the Deltaspike project, DES, DV8, and S101 report 2, 29, and 9 sets of package cycles, involving 8, 34, and 44 packages respectively. The following three columns list the number of packages shared by each pair of tools: there were 7 packages shared by DV8 and DES, 8 shared by DES and S101, and 34 shared by DV8 and S101. It is clear that the 34 cycle-involving packages detected by DV8 either are a subset of the 44 of S101, or have significant overlaps. But the number of cycle-involving packages reported by DES are always significantly less. 

At the file level, DES only reports that a file is involved in a cycle, but does not report how many cycles are detected, or which files are in which cycle. Thus we consider the number of file cycles detected by DES as 1.  For example, in Deltaspike, the cell in the first row, ``\emph{File Cycle Comparison}" section, contains ``$1-9-12$", meaning that DV8 detected 9 cycles, and S101 detected 12. The total number of files involved in cycles are 15, 33, and 45 respectively detected by DES, DV8, and S101. The last three columns list the number of cycle-involving files shared by each pair of tools. 

We observed that while the cycle-involving files reported by S101 and DV8 overlap, the differences are not trivial. In HBase, S101 reported 656 cycle-involving files while DV8 reported 521, with a difference of 135. In some projects, like Flume, even though the number of files reported are similar, the reported cycles they participate in are different, which we will elaborate on in the next section. For all 10 projects, DES only detects about half the number of cycle-involving files with respect to DV8 and S101.

\textbf{Answer to RQ1:} The results, especially after normalizing by size, revealed that there is little agreement between the tools regarding which are the most problematic files, particularly when considering their `flagship' measures, such as smells in SonarQube and Designite, XS in S101, and anti-patterns in DV8.  Moreover, these tools disagree on the most basic measures such as complexity, file cycles and package cycles. 

This raises two questions that we will discuss next: why are these tools report such different results, even though their definitions are similar?  And, most important, which measures are more likely to indicate truly problematic files---those that incur high maintenance costs?

\subsection{RQ2: Where does the disagreement come from?}
\label{sec:q2}
In this section, we analyze in depth why these tools cannot even agree on the most commonly used measures including size, complexity, file cycles, package cycles, as well as smells. 

\textbf{1. Size calculation.} As mentioned earlier, we did not initially realize (or expect) that for the same file, DV8, Designite and S101 report different size numbers (in LOC).  
These differences could be huge. For $ProtosFactory.java$, the three tools reported 5991, 6301, and 13533 LOC respectively. As the most fundamental measure, such differences in LOC are too huge to be acceptable. As shown in Table \ref{tbl:q2-size-pairs}, the average number of top 20 overlaps between DES and S101 is only 15.7. 

We thus used SCC to independently verify which number is most reliable. SCC reports two size numbers: SCC\_LOC--the total number of lines of text, and SCC\_CLOC--the number of source code lines. We observed that the SCC\_CLOC count and DV8\_LOC are almost exactly the same. We only noted a few cases where these numbers differ by 1 LOC. The SCC\_LOC count, however, does not match either S101 or DES size counts. For example, for $ProtosFactory.java$ in Flume, the SCC\_LOC count is 7631, meaning that the lines of code reported by S101 (13533) is larger than the physical lines of the file. We cannot explain why. We chose to use  SCC\_LOC count as the basis for normalization.

\textbf{2. Complexity calculation.} Table~\ref{tbl:q2-complexity-pairs} and \ref{tbl:q2-norm-complexity-pairs} have shown that although these measures are all based on cyclomatic complexity, their correlations can be as low as 0.56, and the highest average cross-tool top-20-overlaps is only 13.5, meaning that many of the most complex files recognized by one tool may not be considered as most complex by another. 

After a deeper look at the tool websites and data, we realized that the differences first come from the calculation of cyclomatic complexity for each function, and second come from how complexity at file-level is calculated. In SCC, function complexity is calculated by counting the number of branches: \textit{``This is done by checking for branching operations in the code. For example, each of the following $for, if, switch, ...$ encountered in Java would increment that files complexity by one."} There are many files for which SCC complexity are 0, meaning that there are no branches in the file. However, according to its definition~\cite{mccabe:tse1976}, cyclomatic complexity should be ``\emph{the number of linearly independent paths}", and the minimal score should be one for a function without any branches.

For DES complexity, even  though all the scores are larger than 1, after examining several files, their complexity score doesn't match the original definition either.  S101\_fat at the method level is defined as the difference between the measured complexity and a given threshold, which makes the outcome different from that of DES and SCC. 

When file-level complexity is considered, these tools deviate even more. For DES,  file-level complexity is calculated as: \textit{``Weighted Methods per Class (WMC) ... the sum of cyclomatic complexities of all the methods belonging to the class. "} While in S101, file-level S101\_fat is defined as the number of dependencies among inner methods, basically have little to do with cyclomatic complexity. For SCC, it is not clear how file-level complexity was calculated, but the scores are more close to that of DES. 

In summary, these tools provide different complexity scores both due to the different implementation of cyclomatic complexity, and how method-level complexity scores are aggregated into file-level scores, which again, raises the question: which one should developers trust? 



\textbf{3. Cycle calculation.} Table~\ref{tbl:q2-all-cycles} reveals that, not only are the number of package cycle groups different, but these tools also disagree on whether a package participated in a cycle or not. The differences between the number of groups is understandably due to the definitions of cycles: in DV8, package cycle is defined as pair-wise dependencies among packages, while the other two tools define a package cycle as a strongly connected graph, which explains why DV8 always reports the largest number of sets. For S101 and DES, even though their package-cycle definitions are similar, the detected number of sets are often different except for HBase (10 sets) and Qpid-B (8 sets).

Even for those projects where the same number of sets are reported, the specific packages that participate in these cycles are vastly different. For example, in HBase, although both DES and S101 reported 10 package cycles, the participating members are 52 (DES) vs. 85 (S101), while the number reported by DV8 is 71. In NiFi, the number of cycle-involving packages was reported to be 41, 157 and 164 by the three tools. 



We observed similar discrepancies from file-cycle data. In terms of the number of cycle-involving files, DV8 is close to S101, but DES reported about 50\% fewer files. For example, in DeltaSpike, DES, DV8 and S101 reported 15, 33, and 45 cycle-involving files respectively. DES and DV8 reported 10 files in common, meaning that there are 23 files that DV8 detected to have cycles, but DES did not identify. 

\begin{figure}
  \centering
  \begin{subfigure}[b]{0.35\textwidth}
        \includegraphics[width=\textwidth]{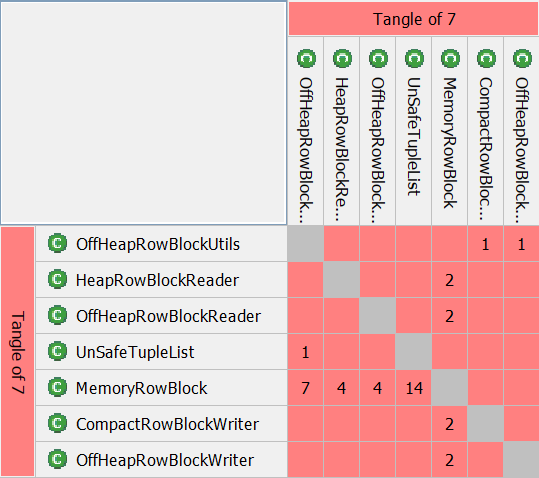}
        \caption{A Tangle Detected by S101}
        \label{fig:s101-tangle}
     \end{subfigure}
  \hfill
  \begin{subfigure}[b]{0.4\textwidth}
    \includegraphics[width=\textwidth]{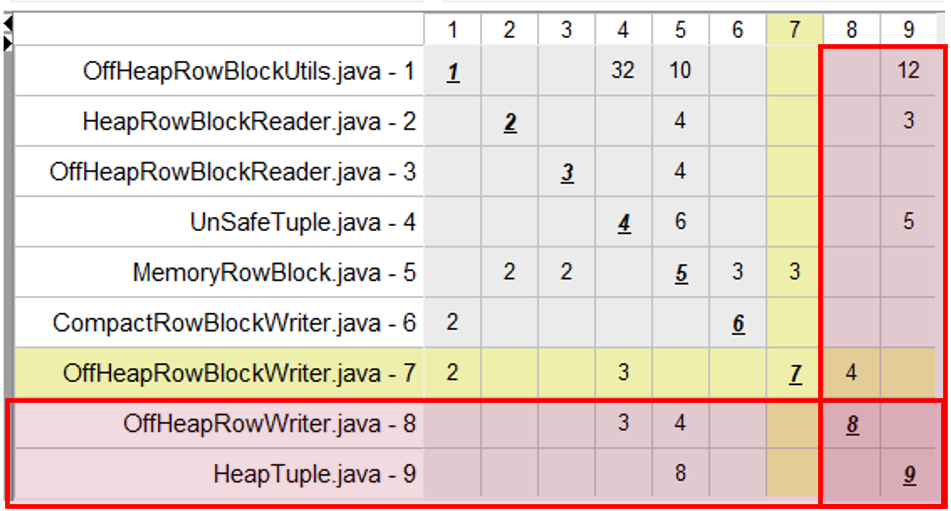}
    \caption{A Clique Detected by DV8}
    \centering
    \footnotesize{(Containing the 7 files in S101 tangle shown above)}
    \label{fig:dv8-clique}
  \end{subfigure}
  \caption{File Cycle Detected by S101 and DV8}
  \vspace{-0.8cm}
\end{figure}

Since Designite Java does not provide a GUI for us to examine where the differences come from, nor does it export information about which files participated in which cycle, we cannot explain the difference between DES and the other two tools. For S101 and DV8, we first noticed that all the cycles in S101 are mutually exclusive; that is, a file can only belong to one cycle. In DV8, a file can belong to multiple cycles. We studied several cycles in both tools, and observed the cases where files should be belong to a cycle as detected by DV8, but are left out by S101. For example, the 7-file tangle detected by S101, as shown in Figure~\ref{fig:s101-tangle}, are contained in the 9-file clique detected by DV8 as shown in Figure~\ref{fig:dv8-clique}. It is clear that the other two files, $OffHeapRowWriter.java$ and $HeapTuple.java$ are also part of the connected graph. It is not clear why they are left out in the S101 tangle. We also notice that the number and types of dependencies between files are calculated differently by S101 and DV8. 

In summary, even though all these tools detect cycles as strongly connected components, the specific instances of cycles and the members within each cycle are calculated differently.

\textbf{4. Smell detection.}
Given that these tools differ in their detection of most commonly used measures, it is not unexpected that when these tools define their own ``smells", the most problematic files recognized by each tool are drastically different. DES defines three categories of smells: implementation (code), design (file), and architecture (package) smells, each having 11, 20, and 7 types.  Structure 101 features tangles and excessive complexity (XS), DV8 defines 6 anti-patterns, and SonarQube allows the user to customize the definitions of smells. By the time we used SonarQube to analyze these subjects, there are 377 rule defined for Java code smells. 

\textbf{Answer to RQ2:} 
Surprisingly there is substantial disagreement between the tools in measures as basic as size and complexity. Some discrepancies are due to differing implementations of the same concept, or the lack of unified definitions of the same term. In most cases, however, finding clear explanations on how the measures are calculated is not straightforward, and the discrepancies are hard to explain.

\subsection{RQ3: Which tools indicate true debt?}

\noindent
Given all these discrepancies reported by these tools, now we turn to the question about which measures are more likely to correctly identify source files with significant technical debt. 

As mentioned previously, for each file in each project we collected four measures,  \emph{\#Change} (\#C), \emph{\#ChangeChurn} (\#CC), \emph{\#Bug} (\#B), and \emph{\#BugChurn} (\#BC) to approximate maintenance costs. These approximate  the \textit{interest} incurred by technical debt. DV8 calculates the measures automatically from the revision history (with respect to the provided date range). As these numbers will act as the benchmark of our comparison, we created an independent tool called JEC to validate them, and confirmed their accuracy. 

\begin{table}[htbp]
\centering
\caption{\#Projects where Measures Agree}
\centering\footnotesize{File Measures vs. Maintenance Measures}
\footnotesize
\begin{center}
\begin{tabular}{|l|c|c|c|c||c|c|c|c|} 
\hline
   \multirow{2}{*}{Measure Name} & \multicolumn{4}{c||}{Correlation $\geq$ 0.5} & \multicolumn{4}{c|}{\#Top-20-Overlaps $\geq$ 10} \\ 
   \cline{2-9} 
 &\#B & \#BC &\#C & \#CC & \#B & \#BC & \#C & \#CC \\ \hline
DV8\_size &\cellcolor{lightgray}7 &\cellcolor{lightgray}6 &\cellcolor{lightgray}7 &\cellcolor{lightgray}5 &3 &2 &4 &2\\ \hline
DES\_size &\cellcolor{lightgray}6 &\cellcolor{lightgray}5 &\cellcolor{lightgray}7 &4 &2 &2 &3 &2\\ \hline
S101\_size &\cellcolor{lightgray}6 &\cellcolor{lightgray}5 &\cellcolor{lightgray}7 &\cellcolor{lightgray}5 &3 &2 &3 &1\\ \hline
SCC\_size &\cellcolor{lightgray}7 &\cellcolor{lightgray}6 &\cellcolor{lightgray}7 &\cellcolor{lightgray}5 &3 &2 &4 &2\\ \hline
DES\_comp. &\cellcolor{lightgray}7 &3 &\cellcolor{lightgray}7 &4 &2 &2 &3 &1\\ \hline
SCC\_comp. &\cellcolor{lightgray}6 &4 &\cellcolor{lightgray}7 &4 &3 &2 &3 &1\\ \hline
S101\_fat &4 &1 &\cellcolor{lightgray}6 &1 &2 &2 &2 &1\\ \hline
S101\_xs &3 &2 &3 &2 &2 &2 &2 &1\\ \hline
ARCH\_fanOut &3 &1 &\cellcolor{lightgray}5 & & & &2 &\\ \hline
ARCH\_deps &1 & &1 & & & & &\\ \hline
ARCH\_coCh &\cellcolor{lightgray}5 &1 &\cellcolor{lightgray}7 & & & &1 &\\ \hline
DV8\_issues &\cellcolor{lightgray}6 &1 &\cellcolor{lightgray}9 & &3 &1 &\cellcolor{lightgray}5 &1\\ \hline
DV8\_crossing &\cellcolor{lightgray}5 &2 &\cellcolor{lightgray}6 &1 &2 & &1 &\\ \hline
DV8\_MV &\cellcolor{lightgray}6 &1 &\cellcolor{lightgray}10 &1 &2 &1 &\cellcolor{lightgray}5 &\\ \hline
DV8\_UH &4 &1 &4 & &2 & &2 &\\ \hline
DV8\_UI &4 & &\cellcolor{lightgray}5 & & & & &\\ \hline
SQ\_issues &4 &2 &\cellcolor{lightgray}5 &2 &2 &1 &3 &\\ \hline
\end{tabular}
\end{center}
\label{tbl:q3-with-cost-corr}
\vspace{-0.3cm}
\end{table}

Table~\ref{tbl:q3-with-cost-corr} shows the number of projects in which a file measure has 0.5 or higher correlation with one of the four maintenance cost measures. The table shows that for all 10 projects, size and complexity measures, as well as ARCH\_fanOut, ARCH\_coCh, DV8\_issues, DV8\_crossing, DV8\_MV, DV8\_UI and SQ\_issues were positively correlated with at least one maintenance measure in 5 or more projects. In particular, DV8\_MV correlates with \emph{\#Changes} in all 10 projects, performing better than all size and complexity measures. 

However, after normalization, as shown in Table~\ref{tbl:q3-norm-with-cost-corr}, only DV8\_MV appears to be correlated with maintenance measures in 5 projects. The implication is that modularity violation appears to be a distinctive measure, more likely to locate technical debt above and beyond what is discernible simply by considering file size. 
\begin{table}[h]
\caption{\#Projects where Normalized Measures Agree}
\centering\footnotesize{Normalized File Measures vs. Maintenance Measures}
\begin{center}
\begin{tabular}{|l|c|c|c|c||c|c|c|c|} 
\hline
    \multirow{2}{*}{Measure Name}  & \multicolumn{4}{c||}{Correlation $\geq$ 0.5} & \multicolumn{4}{c|}{\#Top-20-Overlaps $\geq$ 10} 
    \\ \cline{2-9} 
 &\#B & \#BC &\#C & \#CC &\#B & \#BC &\#C & \#CC \\ \hline
DV8\_issues &1 & &1 & & & & &\\ \hline
DV8\_crossing & & &1 & & & & &\\ \hline
DV8\_MV &4 &1 &\cellcolor{lightgray}5 &1 &2 & & &\\ \hline
DV8\_UI &1 & &2 &1 &1 & &1 &\\ \hline
\end{tabular}
\end{center}
\label{tbl:q3-norm-with-cost-corr}
\end{table}

Considering the top 20 overlaps, as shown in the last 4 columns of  Table~\ref{tbl:q3-with-cost-corr}, it appears that in most cases, the most error-prone or change-prone files are often not the top files ranked by these measures. Only DV8\_issues and DV8\_MV agreed on 10 or more files as the most change-prone in 5 projects. After normalization, as shown in Table~\ref{tbl:q3-norm-with-cost-corr}, only DV8\_MV and DV8\_UI are left in the last 4 columns of the table. It is worth noting that even though ARCH\_coCh also counts the number of co-changes in revision history, it does not perform as well as DV8\_MV, which counts the discrepancy between structure dependencies and co-changes.

\textbf{Answer to RQ3:} These results show that, except for a few measures that incorporate history information, most of the measures calculated by these tools are not more indicative than size and complexity in terms of identifying the most error-prone and change-prone files.  The anti-patterns detected by DV8, especially Modularity Violation and Unstable interface, do provide a distinctive ability to identify debt-ridden files.






\section{Implications}

Our results have enormous implications for practice because if a developer selects and runs one of these tools to analyze their project, they will potentially focus their attention on the wrong set of files.  Specifically:

\begin{itemize}
  \item  Most of the technical debt tools we studied do not give better insight than simply measuring the size of a file.  It has been well established for decades that bigger files generally carry higher risk and require greater cost and effort to maintain. So the cost of these tools, and the effort in using them and interpreting their data, would be largely misguided.  
    \item The tools we studied often disagree with each other, even on basic, seemingly objective measures such as size and complexity. Again, this is problematic for a practitioner as this means that the outputs of some of these tools are unreliable and will lead them to misapply their effort.
    \item Many of the tool measures that we studied have low correlations with our outcome measures of technical debt: bugs, changes, and churn. Since a practitioner using a tool would be highly motivated to manage the effort associated with revisions bugs, changes, and churn, these tool measures are giving little insight into how to do this and where a practitioner should apply their effort.  
    \item Many tools report file and package cycles, and explicitly recommend that project effort should be devoted to breaking these cycles. But not all the cycles reported are accurate, and these measures turn out not to be strong predictors, across projects, of future maintenance cost and complexity. 
\end{itemize}

There is, however, one ray of hope.  Measures that take into account co-change (historical) information do perform significantly better than purely code-based (structural) measures. Thus projects should spend the effort to link commits to issues, to obtain deeper understanding of which  debts influence their maintenance costs, and how. Overall, the results of our study suggest that using an analysis tool that does not consider change history will result in sub-optimal debt remediation efforts.




\section{Threats to validity and future work}
\label{sec:discussion}
In this section, we discuss the threats to validity of our empirical study, and our plans for future work. 

\textit{Threats to Validity.}
The major threats to the conclusion of this study include the subjects we chose, the targeted tools, and using revisions and churn as the ground truth of maintainability. Our study was conducted only for programs written in Java. It may be possible (but, we believe, unlikely) that results may differ for other languages.

In this paper, we only chose six tools for the reasons explained in Section \ref{sec:empiricalstudy}. Many other tools also detect technical debt in various forms, such as  CAST~\cite{castsoftware},  SonarGraph~\cite{sonargraph},  CodeInspector~\cite{codeinspector},  CodeMRI~\cite{codemri}, SQuORE~\cite{squore},  SymfonyInsight~\cite{symfony}, inFusion~\cite{thanis:jserd2017}, JDeodorant~\cite{tsantalis:csmr08}, PMD~\cite{pmd}, and Checkstyle~\cite{checkstyle}. In addition to Archinaut and DV8, CodeScene\cite{codescene:codescene}
also analyze revision history, but we cannot export results from it. Archinaut calculates some similar measures to it. 

It is not clear if and to what extent other tools will locate technical debt differently than the tools we studied. But this work reveals the lack of consistency and unified definitions among several widely used technical debt detection tools.

Using revisions and churn as ground truth for maintainability measures could be too simplistic. For projects that are not being actively maintained, these numbers will not be able to accurately reflect technical debt. It is also possible that a system has too much technical debt, cannot evolve any more, and thus has very low churn or revisions.  In this case, this study may be biased showing the files have little technical debt. To mitigate this threat, we have chosen an analysis period not based on time but rather based on the same number of commits that are linked to issues. 

It is arguable that measures containing co-change information, such as Modularity Violation, perform better in terms of identifying high-maintenance files because of construction bias since they already considered the number of revisions. However, our study showed that the measure that only counts co-changes does not perform better than size and complexity, but DV8\_MV does. It also correlates with the number of bug revisions, which is not part of the definition. 

\textit{Future Work.} In addition to addressing these threats to validity, this initial empirical study pointed to several interesting directions of research. First, we would like to further evaluate if and to what extent the overall scores provided by each tool are consistent with each other. We have observed that SonarQube rates all 10 projects ``grade A'' (the lowest level of technical debt), but the other tools give rather different measures of the quality of these projects. For example DV8's decoupling level metric~\cite{mo:icse2016}--which measures how well a project is decoupled and modularized--ranges from 52.35\% to 93.9\% for the 10 projects, suggesting an enormous range in their maintainability.  


Given that there are many tools that purport to detect technical debt, and given the lack of consistency among their definitions and results, this study suggests a need. We need to create a technical debt benchmark and a widely validated metrics suite so that  projects can confidently assess, track, and compare their quality between projects and over time. The creation of such a benchmark would require  a more rigorous model of file-level technical debt.

\section{Related Work}
\label{sec:background}
To the best of our knowledge, Fontana et al.~\cite{fontana:jot2012} was the first to present a comparative evaluation of code smell detection tools. They used six versions of a system to evaluate four tools, Checkstyle, inFusion, JDeodorant, and PMD.  Fernandes et al. \cite{Fernandes:ease2016} also reported a systematic literature review on 29 tools that were evaluated considering two smells, Large Class and Long Method. They studied two software systems and three metrics for comparison: agreement, recall, and precision. Thanis et al.~\cite{thanis:jserd2017} calculated the accuracy of  4 tools in the detection of three code smells: God Class, God Method, and Feature Envy. They calculated the agreement between pairs of tools. One of their  findings is that the tools have different accuracy levels under different contexts. For example, the average recall of a project can vary from 0 to 58\% and the average precision can vary from 0 to 100\%.  
These prior studies focus on \emph{code} smells only at the level of methods. From a design and architectural perspective, we are more interested to know which files contain true debt, that is, files associated with excessive maintenance effort?

Avgeriou et al. ~\cite{Avgeriou:ieee2020} recently published a comparison study of 9 technical debt measurement Tools, including DV8 and SonarQube. This work compared these tools in terms of their popularity and features, such as the number of related publications. By contrast, our study analyzed 10 projects and compared their analysis outputs in detail. 

The study reported in this paper, using 6 tools and 10 open source project is, to the best of our knowledge,  the first comparative study of technical debt at the file level taking maintenance effort into consideration.

\section{Conclusion}
In this paper, we reported on our empirical study of 6 file-level technical debt detection tools applied to 10 open source projects. Our objective was to investigate if and to what extent these tools agree on which files contain technical debt, what is the root cause of inconsistency, and which tools and measures are more likely to pinpoint files associated with high maintenance effort. We calculated pair-wise measurement correlations and intersections between the measures, as well as with churn and revision counts for each file that we studied. 

Our results clearly show that without co-change (historical) information, purely code-based (structural) measures do not agree with each other in terms of which files contain technical debt, other than trivially identifying large files, which could just as easily have been identified by counting LOC. It is a surprise that these tools disagree on even the most commonly used measures of size, complexity, and cycles. Moreover, these code-based measures, after normalizing by LOC, do not identify files that incur the most maintenance costs in practice. In other words, in terms of identifying the most problematic files, our data suggests that some of the best known code-based tools are not producing more insight than ``\emph{big files are bad}". 

Measures using both structural and historical information perform better in terms of correlating with high-maintenance files, and provide additional insights into the root causes of the debt, and hence these measures have the potential to guide refactoring. For example, the Modularity Violation and Unstable Interface anti-patterns that DV8 identifies give insight into not only what files are problematic, but also how to refactor those files to remove the root causes of the debt. This study highlights the necessity of leveraging historical information in technical debt detection, and the need to build a more rigorous, uniform model of technical debt, as well as validated benchmarks. 




\section{Data Availability}
Our data is available at \href{https://zenodo.org/record/4588039}{https://zenodo.org/record/4588039}.

\section{Acknowledgments}
This work was supported in part by the National Science Foundation under grants CCF-1816594/1817267, OAC-1835292, and CNS-1823177/1823214. 

\bibliographystyle{IEEEtran}
\bibliography{references}

\end{document}